\documentclass[12pt]{article}
\usepackage{array}
\usepackage{epsfig}

\usepackage{amssymb}

\makeatletter
\def\alt{\mathrel{\mathpalette\gl@align<}}
\def\agt{\mathrel{\mathpalette\gl@align>}}
\def\gl@align#1#2{\lower.6ex\vbox{\baselineskip\z@skip\lineskip\z@
\ialign{$\m@th#1\hfil##\hfil$\crcr#2\crcr\sim\crcr}}} \makeatother

\def\bwt{\begin{widetext}}
\def\ewt{\end{widetext}}
\def\be{\begin{equation}}
\def\ee{\end{equation}}
\def\bea{\begin{eqnarray}}
\def\eea{\end{eqnarray}}
\def\bean{\begin{eqnarray*}}
\def\eean{\end{eqnarray*}}
\def\bary{\begin{array}}
\def\eary{\end{array}}
\def\bit{\begin{itemize}}
\def\eit{\end{itemize}}

\def\su5u1{SU(5) \times U(1)}
\def\fsu5u1{SU(5) \times U(1)'}
\def\so10{SO(10)}
\def\sq20{SO(10) \times SO(10)}

\begin{document}
\begin{flushright}
{\tt hep-ph/0602040}\\
BA-06-01\\
\end{flushright}
\vspace*{1.0cm}
\begin{center}
{\baselineskip 25pt \Large{\bf
Higgs Boson Mass From Orbifold GUTs \\[2.5mm]
}}

\vspace{1cm}

{\large Ilia Gogoladze$^a$\footnote {On a leave of absence from:
Andronikashvili Institute of Physics, GAS, 380077 Tbilisi, Georgia.
\\ \hspace*{0.5cm} email: {\tt ilia@physics.udel.edu}}, Tianjun Li$^{b,c}$\footnote
{email: {\tt tjli@physics.rutgers.edu}}  and Qaisar
Shafi$^d$\footnote {email: {\tt shafi@bartol.udel.edu}} }
\vspace{.5cm}

{\small {\it $^a$Department of Physics and Astronomy, University of
Delaware, Newark, DE 19716, USA \\
$^b$Department of Physics and Astronomy, Rutgers University,
Piscataway, NJ 08854, USA\\
$^c$Institute of Theoretical Physics, Chinese Academy of Sciences,
 Beijing 100080, P. R. China \\
$^d$ Bartol Research Institute, Department of Physics and Astronomy,
University of Delaware, Newark, DE 19716, USA }}


\date{\today}

\vspace{1.0cm} {\bf Abstract}

\end{center}

\baselineskip 16pt

We consider a class of seven-dimensional ${\cal N}=1$ supersymmetric
orbifold GUTs in which the Standard Model (SM) gauge couplings and one of
the Yukawa couplings (top quark, bottom quark or tau lepton) are
unified, without low energy supersymmetry, at $M_{\rm GUT} \simeq
4\times 10^{16}$ GeV. With gauge-top quark Yukawa coupling
unification the SM Higgs boson mass is estimated to be $135\pm  6$
GeV, which increases to $144\pm 4$ GeV for gauge-bottom quark (or
gauge-tau lepton) Yukawa coupling unification.

\thispagestyle{empty}

\bigskip
\newpage

\addtocounter{page}{-1}

\section{Introduction}

It was recently shown that the Standard Model (SM) gauge couplings
can be unified at a scale $M_{\rm GUT} \sim 10^{16}-10^{17}$ GeV
provided one employs a non-canonical $U(1)_Y$
normalization~\cite{Barger:2005gn}. This can be realized, for
instance, within the framework of suitable higher-dimensional
orbifold grand unified theories (GUTs)~\cite{Orbifold,Li:2001tx} in
which the scale of supersymmetry breaking, via the Scherk-Schwarz
mechanism~\cite{Scherk:1978ta}, is assumed to be comparable to
$M_{\rm GUT}$. Such a high scale of supersymmetry breaking is partly
inspired by the string landscape~\cite{String}. The SM Higgs field
in this case is identified with an internal component of the  gauge
field. For some recent papers on   gauge--Higgs unification see Ref.
\cite {Antoniadis:2001cv}. The SM Higgs mass in a class of
seven-dimensional (7D) orbifold GUTs was estimated to lie in the
mass range of 127--165 GeV ~\cite{Barger:2005gn}.

 In this paper we take the orbifold GUTs in Ref.~\cite{Barger:2005gn}
 a step  further by
including a new ingredient. We consider compactification schemes in
which the gauge coupling unification is extended to also include one
of the Yukawa couplings from the third family.  Thus, by unifying
the top quark Yukawa coupling at $M_{\rm GUT}$ with the three SM
gauge couplings, we are able to provide a reasonably  precise
estimate for the SM Higgs mass, namely $135 \pm 6$ GeV.
 Replacing the top quark Yukawa coupling
with the bottom quark or tau lepton Yukawa coupling  leads to a
somewhat larger value of the Higgs mass ($144\pm 4$ GeV).  Note that
the  gauge--Yukawa coupling unification in orbifold GUTs was
investigated earlier within  low-scale supersymmetry in
Ref.~\cite{GY}.

The plan of this paper is as follows. In Section 2 we briefly
summarize the 7D $SU(7)$ orbifold model (with some technical details
in Appendix A). Section 3 is devoted to the unification of gauge and
top quark Yukawa coupling. Figure 1 displays the unification scale
as well as the magnitude of the unified coupling. Figure 2 shows a
plot of the Higgs mass versus the top quark mass $m_{\rm top}$. For
the current central value $m_{\rm top} = 172.7$
GeV~\cite{Beneke:2000hk}, the corresponding Higgs mass is close to
135 GeV. In Sections 4 and 5 we replace the top quark Yukawa
coupling with the bottom quark and tau lepton Yukawa couplings,
respectively. The results for the bottom quark case are displayed in
Figs. 3 and 4. The Higgs mass turns out to be somewhat larger than
for the top quark case, with a central value close to 144 GeV. The
tau lepton case is very similar to the bottom quark case. In Section
6 we consider a 7D $SU(8)$ model in which the SM gauge couplings and
the top and bottom quark Yukawa couplings are all unified at $M_{\rm
GUT}$ (A scenario of this kind with low-energy
 supersymmetry has previously been discussed in~\cite{GY}).
Our conclusions are summarized in Section 7.

\section{$SU(7)$ Orbifold Models}

To realize gauge--Yukawa unification we consider a 7D ${\cal
N}=1$ supersymmetric  $SU(7)$ gauge theory compactified on the
orbifold $M^4\times T^2/Z_6 \times S^1/Z_2$ (for some details see
Appendix A). We find that $SU(7)$ is the smallest gauge group which
allows us to implement gauge--Yukawa unification at $M_{\rm GUT}$
with a non-canonical normalization $k_Y = 4/3$ for $U(1)_Y$. The
${\cal N}=1$ supersymmetry in 7D has 16 supercharges corresponding
to ${\cal N}=4$ supersymmetry in 4-dimension (4D), and only the
gauge supermultiplet can be introduced in the bulk.  This multiplet
can be decomposed under  4D
 ${\cal N}=1$ supersymmetry into a gauge vector
multiplet $V$ and three chiral multiplets $\Sigma_1$, $\Sigma_2$,
and $\Sigma_3$  all in the adjoint representation, where the fifth
and sixth components of the gauge field, $A_5$ and $A_6$, are
contained in the lowest component of $\Sigma_1$, and the seventh
component of the gauge field $A_7$ is contained in the lowest
component of $\Sigma_2$. As pointed out in Ref.~\cite{NMASWS}
the bulk action in the Wess-Zumino gauge and in 4D ${\cal N}=1$
supersymmetry notation contains trilinear terms involving  the
chiral  multiplets $\Sigma_i$. Appropriate choice of the orbifold
enables us to identify some of them with the SM Yukawa couplings~\cite{GY}.

To break the $SU(7)$ gauge symmetry, we select the following
$7\times 7$ matrix representations for $R_{\Gamma_T}$ and
$R_{\Gamma_S}$ defined in Appendix A
\begin{eqnarray}\label{bb1}
R_{\Gamma_T} &=& {\rm diag} \left(+1, +1, +1,
 \omega^{n_1}, \omega^{n_1}, \omega^{n_1}, \omega^{n_2} \right),
 \label{bb3}
\end{eqnarray}
\begin{eqnarray}\label{bb4}
 R_{\Gamma_S} &=& {\rm diag} \left(+1, +1, +1, +1, +1,
-1, -1 \right),
\end{eqnarray}
where $n_1$ and $n_2$ are positive integers, and $n_1 \not= n_2$.
 Then, we obtain
\begin{eqnarray}
&& \{SU(7)/R_{\Gamma_T}\} ~=~ SU(3)_C\times SU(3)\times U(1) \times
U(1)^{\prime},
\nonumber \\
 &&\{SU(7)/R_{\Gamma_S}\} ~=~ SU(5)\times SU(2) \times
U(1),
\end{eqnarray}
\begin{eqnarray}
 \{SU(7)/\{R_{\Gamma_T} \cup R_{\Gamma_S}\}\}
~=~ SU(3)_C\times SU(2)_L\times U(1)_Y \times U(1)_{\alpha} \times
U(1)_{\beta}. \label{bb2}
\end{eqnarray}
So, the 7D ${\cal N}= 1 $
supersymmetric gauge symmetry $SU(7)$ is broken down to  4D
${\cal N}=1$ supersymmetric gauge symmetry $SU(3)_C\times
SU(2)_L\times U(1)_Y \times U(1)_{\alpha} \times U(1)_{\beta} $~\cite{Li:2001tx}.
In Eq. (\ref{bb2}) we see the appearance of two U(1) gauge
symmetries which we assume can be spontaneously broken at or close
to $M_{\rm GUT}$ by the usual Higgs mechanism. It is conceivable
that these two symmetries can play some useful role as flavor
symmetries \cite{FN}, but we will not pursue this any further here.
A judicious choice of $n_1$ and $n_2$ will enable us to obtain the
desired zero modes from the multiplets  $\Sigma_i$ defined in
Appendix A.

The $SU(7)$ adjoint representation $\mathbf{48}$ is decomposed under
the $SU(3)_C\times SU(2)_L\times U(1)_Y \times U(1)_{\alpha} \times
U(1)_{\beta}$ gauge symmetry as:
\begin{equation}
\mathbf{48} = \left(
\begin{array}{cccc}
\mathbf{(8,1)}_{Q00} & \mathbf{(3, \bar 2)}_{Q12}
& \mathbf{(3, 1)}_{Q13} & \mathbf{(3,1)}_{Q14} \\
 \mathbf{(\bar 3,  2)}_{Q21} & \mathbf{(1,3)}_{Q00}
& \mathbf{(1, 2)}_{Q23} & \mathbf{(1, 2)}_{Q24} \\
\mathbf{(\bar 3, 1)}_{Q31} & \mathbf{(1, \bar 2)}_{Q32}
& \mathbf{(1, 1)}_{Q00} & \mathbf{(1, 1)}_{Q34} \\
\mathbf{(\bar 3, 1)}_{Q41} & \mathbf{(1, \bar 2)}_{Q42}
& \mathbf{(1, 1)}_{Q43} & \mathbf{(1, 1)}_{Q00}
\end{array}
\right) +  \mathbf{(1,1)}_{Q_{00}}\, ,
\label{48arj}\end{equation} where the  $\mathbf{(1,1)}_{Q00}$ in the
third and fourth diagonal entries of the matrix and the last term
$\mathbf{(1,1)}_{Q_{00}}$ denote the gauge fields
associated with $U(1)_Y \times U(1)_{\alpha} \times U(1)_{\beta} $.
The subscripts $Qij$, which are anti-symmetric ($Qij=-Qji$), are the
charges under  $U(1)_Y \times U(1)_{\alpha} \times U(1)_{\beta}$.
The subscript $Q00~=~(\mathbf{0}, \mathbf{0}, \mathbf{0})$, and the
other subscripts $Qij$ with $i\not= j$ will be given for each model
explicitly.

\section{Unification of Gauge  and Top Quark Yukawa Couplings }

To achieve gauge and top quark Yukawa coupling unification at $M_{\rm
GUT}$, we make the following choice
\begin{eqnarray}
n_1~=~5~~ {\rm and}~~n_2~=~2~~~{\rm or}~~~3\,,~\, \label{N-numberA}
\end{eqnarray}
in Eq. (\ref{bb1}).  This allows  us to obtain zero modes from
$\Sigma_i$ corresponding to the up and down Higgs doublets $H_u$ and
$H_d$, as well as the left- and right-handed top quark superfields.
The SM Higgs field arises, of course, as a linear combination
 of $H_u$ and $H_d$~\cite{Barger:2005gn}.

\renewcommand{\arraystretch}{1.4}
\begin{table}[h]
\begin{center}
\begin{tabular}{|c|c|}
\hline
Chiral Fields & Zero Modes  \\
\hline\hline
$\Sigma_1$ & $Q_3$:~ $\mathbf{(3, \bar 2)}_{Q12}$ \\
\hline $\Sigma_2$ & ~$H_u$:~ $\mathbf{(1, 2)}_{Q23}$;
~$H_d$:~ $\mathbf{(1, \bar 2)}_{Q32}$ \\
\hline
$\Sigma_3$ & $t^c$: $\mathbf{(\bar 3, 1)}_{Q31}$ \\
\hline
\end{tabular}
\end{center}
\vspace{-0.3cm} \caption{\small Zero modes from  the chiral
multiplets $\Sigma_1$, $\Sigma_2$ and $\Sigma_3$  with gauge and top
quark Yukawa coupling unification. \label{Spectrum-UP-7D}}
\end{table}

 The generators for the gauge symmetry $U(1)_Y
\times U(1)_{\alpha} \times U(1)_{\beta}$ are as follows:
\begin{eqnarray}
&&T_{U(1)_{Y}} \equiv {1\over 6} {\rm diag}\left(1, 1, 1, 0, 0, -3,
0 \right) + ~{{\sqrt {14}}\over {42}} {\rm diag}\left(1, 1, 1, 1, 1,
1, -6 \right), \label{SU7-GU1Y} \nonumber \\ && T_{U(1)_{\alpha}}
\equiv -{{\sqrt {14}}\over {2}} ~{\rm diag}\left(1, 1, 1, 0, 0, -3,
0 \right) + ~{\rm diag}\left(1, 1, 1, 1, 1, 1, -6 \right), \nonumber
\\&& T_{U(1)_{\beta}} \equiv {\rm diag}\left(1, 1, 1, -2, -2, 1, 0
\right), \label{SU7-GU1A}
\end{eqnarray}
With a canonical normalization ${\rm tr}[T_i^2]=1/2$ of non-abelian
generators, from Eq. (\ref{SU7-GU1A}) we find ${\rm tr}
[T_{U(1)_{Y}}^2]=2/3$. For  $k_Y g_Y^2 = g_2^2 = g_3^2$
at the GUT scale, this gives
$k_Y = 4/3$. It was shown in~\cite{Barger:2005gn}
 that the two-loop gauge coupling
unification in this case occurs at $M_{\rm GUT} \simeq 4 \times
10^{16}$ GeV. In our following numerical work we will use this
to estimate for $M_{\rm GUT}$.

The charge assignments $Qij$ from Eq. (\ref{48arj}) are as follows:
\begin{eqnarray}
 &&Q12 = \left ( \mathbf{1\over 6}, \mathbf{-{{\sqrt {14}}\over 2}}, \mathbf{3}\right ),~~~~~
 Q14 = \left (\mathbf{{{1+{\sqrt {14}}}\over {6}}},
\mathbf{{{14-{\sqrt {14}}}\over {2}}}, \mathbf{1}\right ),
 \nonumber \\
&& Q13= \left (\mathbf{2\over 3}, \mathbf{-2{\sqrt
{14}}}, \mathbf{0}\right ),
 ~~~~~  Q23 =  \left ( \mathbf{{1\over 2}},
\mathbf{-{{3{\sqrt {14}}}\over {2}}}, \mathbf{-3}\right ),
\nonumber \\
&& Q24=\left ( \mathbf{{{\sqrt {14}}\over 6}},
\mathbf{{7}}, \mathbf{-2}\right
 ), ~~~~~~~ Q34=\left
(\mathbf{{{-3+{\sqrt {14}}}\over {6}}},
\mathbf{{14+3{\sqrt {14}}}\over {2}}, \mathbf{1}\right ).
\end{eqnarray}

Substituting Eq. (\ref{N-numberA}) in Eqs. (\ref{bb3})--(\ref{bb4})
and employing the $Z_6 \times Z_2$ transformation properties Eqs.
(\ref{bb7})--(\ref{bb10}) for the decomposed components of the
chiral multiplets $\Sigma_i$, we obtain the zero modes presented in
Table 1. We can identify them as a pair of Higgs superfields as well as
the left- and right-handed top quark superfields, as desired.

\begin{figure}[]
\centering
\includegraphics[width=8cm]{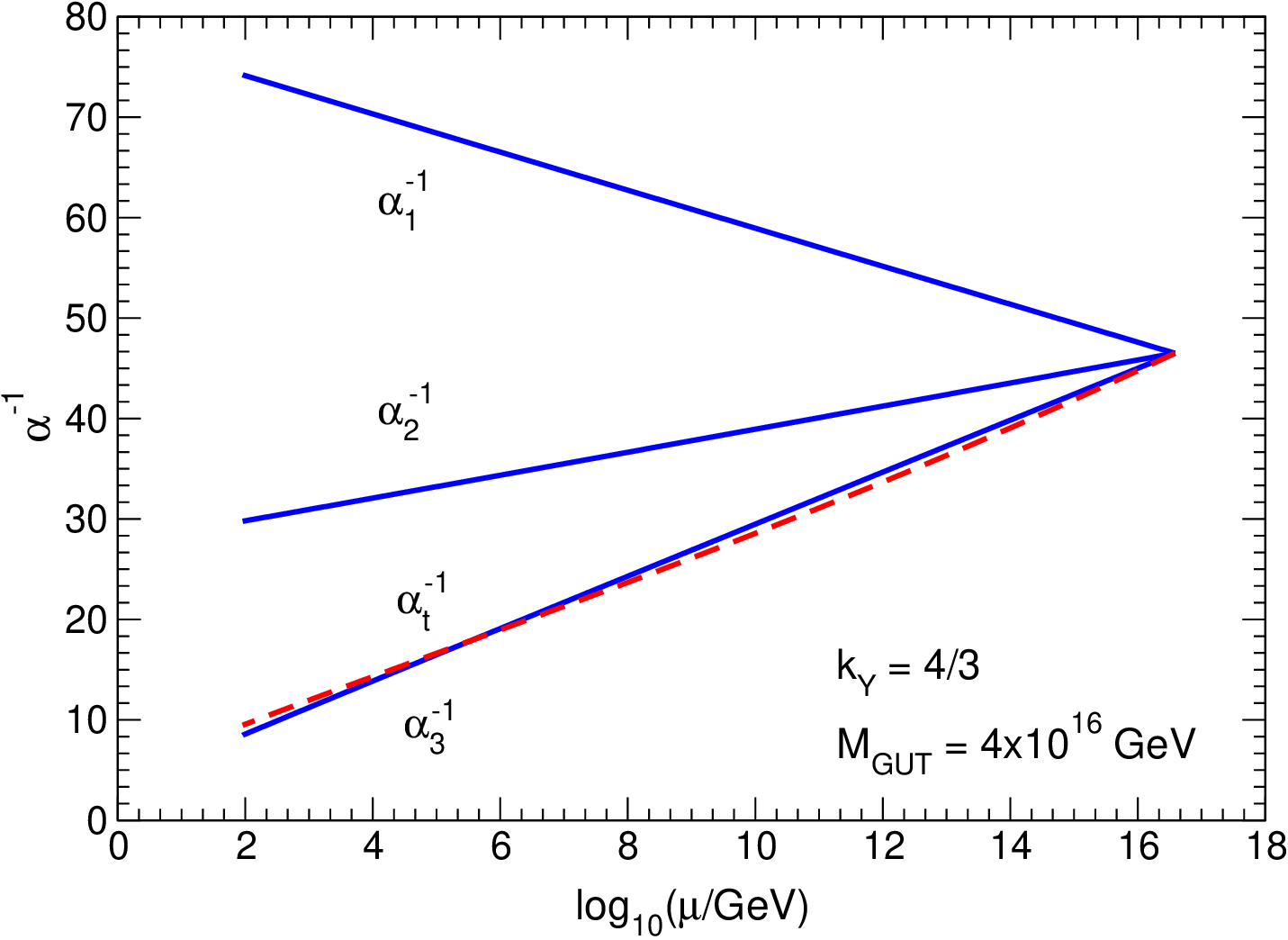}
\vspace{-0.3cm} \caption{\small Two-loop evolution of gauge
 (solid) and  top quark Yukawa (dash) couplings, with
$k_Y=4/3$.}
\end{figure}

From the trilinear term in the 7D bulk action in Eq. (\ref{action7})
the top quark Yukawa coupling is contained in the term
\begin{eqnarray}
\label{bs1}  \int d^7 x \left[ \int d^2 \theta \ g_7 Q_3 t^c H_u
 + h.c.\right],
\end{eqnarray}
where $g_7$ is the $SU(7)$ gauge coupling at the compactification
scale, which  for simplicity,  we identify it as  $M_{\rm GUT}$. Note
that the Higgs superfield $H_u$ appears in Eq. (\ref{bs1}). We
will ignore  brane localized gauge kinetic terms, which may be suppressed
by taking  $VM_{*}\gtrsim O(100)$, where $V$ denotes the volume of
the extra dimensions and $M_{*}$  is the cutoff scale
\cite{Orbifold}. With these caveats we obtain the 4D gauge--top
quark Yukawa coupling
unification at $M_{\rm GUT}$
\begin{eqnarray}
g_1=g_2=g_3=y_t=g_7/\sqrt{V}, \label{un}
\end{eqnarray}
where  $y_t$ is the top quark Yukawa coupling.

The top quark coupling to the SM Higgs will pick up an additional
factor because the latter arises from the linear combination
\begin{eqnarray}\label{hh5}
H\equiv -\cos \beta i \sigma_2\,  H^{*}_d + \sin  \beta \, H_u,
\end{eqnarray}
where $\beta$ is the mixing angle and $\sigma_2$ is the second Pauli
matrix. The effective tree-level top quark Yukawa coupling at $M_{\rm
GUT}$ is then given by
\begin{eqnarray}
h_t=y_t\sin\beta=g_7\sin\beta /\sqrt{V}. \label{un2}
\end{eqnarray}
Note that the linear combination orthogonal to Eq. (\ref{hh5}) is
superheavy and does not play a role in low energy phenomenology. Of
course, the mass scale of $H$ is fine tuned to be of
the order $M_Z$.

One possible way to implement the fine tuning is to introduce a
brane localized gauge singlet field S with a VEV of order $M_{\rm
GUT}$. The superpotential coupling $H_u H_d S$ induces order $M_{\rm
GUT}$ mass terms for the doublets, which combined with order $M_{\rm
GUT}$ supersymmetry breaking soft terms, can yield the desired $M_Z$
scale for $H$ through fine tuning. Note that the Higgsino mass is of
the order $M_{\rm GUT}$, too.

\begin{figure}[]
\centering
\includegraphics[width=9.cm]{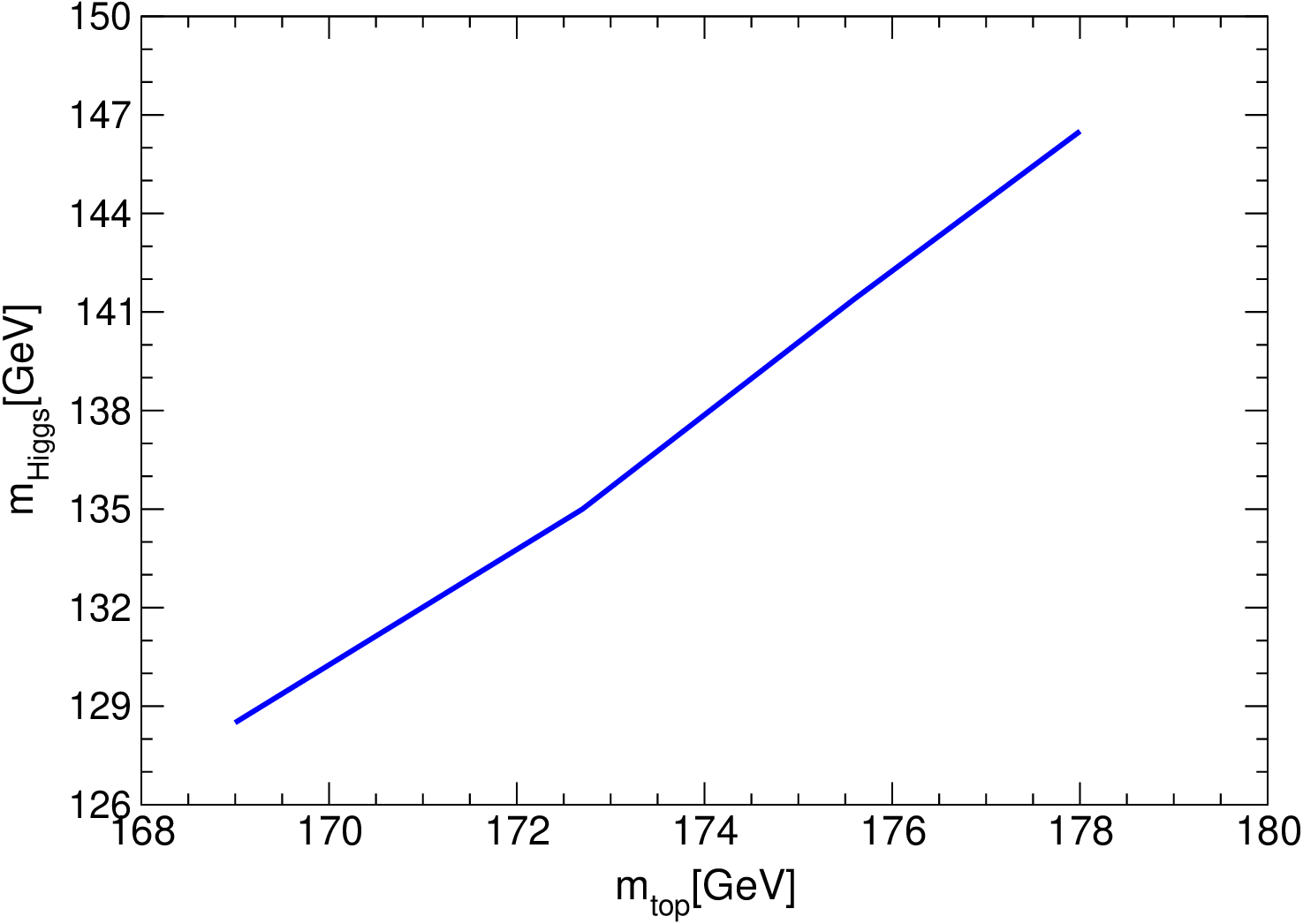}
\vspace{-0.3cm} \caption{\small Higgs boson mass $m_{\rm Higgs}$
versus top quark mass $m_{\rm top}$ with gauge--top quark
Yukawa coupling unification at $M_{\rm GUT}$.}
\end{figure}

The quartic Higgs coupling is determined at $M_{\rm GUT}$ by the
supersymmetric $D$-term
\begin{eqnarray}\label{lam}
\lambda = \frac{\frac{3}{4} g^2_1(M_{\rm GUT})+g^2_2(M_{\rm
GUT})}{4}\cos ^2 2\beta.
\end{eqnarray}
The  renormalization group equation (RGE) for $\lambda$ is given in
Eq. (\ref{ggg7}) in Appendix B. In the numerical calculations we
employ  two-loop RGEs for the gauge, Yukawa couplings, and Higgs
quartic couplings (see Appendix B). There could be threshold
corrections to $\lambda(M_{\rm GUT})$  from the supersymmetric
spectrum, but since we have not specified a scenario for
supersymmetry breaking, we will not consider  them here.

Using $\alpha^{-1}_{EM}(M_Z) = 128.91 \pm
0.02$~~and~~$\sin^2\theta_W(M_Z) = 0.23120 \pm 0.00015$ in
$\overline{MS}$ scheme  \cite{pdg} and   with $k_Y=4/3$,
 we can determine $M_{\rm GUT}$ as well as
 the unified  coupling constant at $M_{\rm GUT}$. Evolving the couplings
from $M_{\rm GUT}$ to  $M_Z$, according to the  boundary condition in
Eq. ({\ref{lam}}), we estimate that $\alpha_3 (m_Z) \simeq 0.118$,
in good agreement with the data \cite{pdg}.

The SM gauge couplings (more precisely $\alpha_i^{-1}$) are plotted
in Fig. 1,  which also displays the coupling $\alpha_t^{-1}\equiv
4\pi/y_t^2$. Knowing  $y_t$  at low energies allows us to estimate
the Higgs  mixing angle $\beta$ in Eq. (\ref{hh5}) by using the
measured value $172.7\pm 2.9$ GeV of the top quark mass
\cite{Beneke:2000hk}. We find $1.3\leq \tan\beta \leq 1.8$,  which
is inserted in Eq. (\ref{lam}) to fix the Higgs quartic coupling
$\lambda(M_{\rm GUT})$.  Employing Eq. (\ref {ggg7}) we can then
determine  $\lambda$  at low energy.

The Higgs boson mass will be estimated by employing the one--loop
effective potential \cite{Altarelli:1994rb}

\be V_{eff} = -m_h^2 H^\dagger H + \frac{\lambda}{2} (H^\dagger H)^2
- \frac{3}{16\pi^2} h_t^4 (H^\dagger H)^2 \left[\log\frac{h_t^2
(H^\dagger H)}{Q^2} - \frac{3}{2}\right], \ee
 where the coefficient
$(-m_h^2)$ of the quadratic term is fine tuned along the line
discussed above. The top quark Yukawa coupling to $H$ is
$h_t=y_t\sin\beta$, and  the scale $Q$ is chosen to coincide with
the Higgs boson mass. In Fig. 2, we plot the Higgs mass versus
$m_{\rm top}$. For the presently favored central value $\rm m_{top}=
172.7\pm 2.9$ GeV \cite{Beneke:2000hk}, we estimate the Higgs mass
to be 135 GeV. It is intriguing that the  Higgs mass estimate is
somewhat higher then the 126 GeV upper bound  on the lightest
neutral  Higgs boson mass in the MSSM \cite{chaber}.

As far as the remaining charged fermions are concerned, we note that
on the 3-brane at the $Z_6\times Z_2$ fixed point
 $(z, y)= (0, 0)$, the preserved gauge symmetry
is $SU(3)_C\times SU(2)_L\times U(1)_Y \times U(1)_{\alpha}\times
U(1)_{\beta}$. Thus, on the observable 3-brane at $(z, y)= (0, 0)$,
we can introduce  the first two
 families of the SM quarks and leptons,
 the right-handed bottom quark,  the $\tau$ lepton doublet,
and the right-handed $\tau$ lepton. The $U(1)_{\alpha}\times
U(1)_{\beta}$ anomalies can be canceled by assigning  suitable
charges to the SM quarks and leptons.  For example,
 under  $U(1)_{\alpha}\times U(1)_{\beta}$ the  charges for
the first-family quark doublet  and the right-handed up quark can be
respectively $(\mathbf{-3}, \mathbf{{\sqrt {14}}/ 2})$ and
$(\mathbf{0}, \mathbf{-2{\sqrt {14}}})$, while the charges of
remaining SM fermions are zero.

\section{Unification of   Gauge and Bottom Quark Yukawa Couplings  }

To implement this scenario we make the following  choice  in Eq.
(\ref{bb1}):
\begin{eqnarray}
n_1~=~5~,~~n_2~=~2~{\rm or}~3~.~\,
\label{bt1}
\end{eqnarray}
The identification of $U(1)_Y$ differs from the  previous  Section.
The generators of  $U(1)_Y \times U(1)_{\alpha} \times U(1)_{\beta}$
are defined as follows:
\begin{eqnarray}
&&T_{U(1)_{Y}} \equiv -{1\over 6}{\rm diag}\left(0, 0, 0, 1, 1, -2,
0 \right) ~+ {{\sqrt {21}}\over {42}}{\rm diag}\left(1, 1, 1, 1,
1, 1, -6 \right) ,~\, \label{SU7-GD1Y} \nonumber \\
&&T_{U(1)_{\alpha}} \equiv {\sqrt {21}} ~ {\rm diag}\left(0, 0, 0,
1, 1, -2, 0
\right)+ {\rm diag}\left(1, 1, 1, 1, 1, 1, -6 \right), \nonumber \\
&&T_{U(1)_{\beta}} \equiv {\rm diag}\left(1, 1, 1, -1, -1, -1, 0
\right). \label{SU7-GD1A}
\end{eqnarray}
Note that $k_Y=4/3$ also in this case.

\renewcommand{\arraystretch}{1.5}
\begin{table}[b]
\begin{center}
\begin{tabular}{|c|c|}
\hline
Chiral Fields & Zero Modes  \\
\hline\hline
$\Sigma_1$ & $Q_3$:~ $\mathbf{(3, \bar 2)}_{Q12}$ \\
\hline $\Sigma_2$ & ~$H_d$:~ $\mathbf{(1, 2)}_{Q23}$;
~$H_u$:~ $\mathbf{(1, \bar 2)}_{Q32}$ \\
\hline
$\Sigma_3$ & $b^c$: $\mathbf{(\bar 3, 1)}_{Q31}$ \\
\hline
\end{tabular}
\end{center}
\vspace{-0.3cm}
 \caption{\small Zero modes from the chiral multiplets
$\Sigma_1$, $\Sigma_2$ and $\Sigma_3$  with gauge  and bottom quark
Yukawa coupling unification. \label{Spectrum-DOWN-7D}}
\end{table}

\begin{figure}[t]
\centering
\includegraphics[width=8cm]{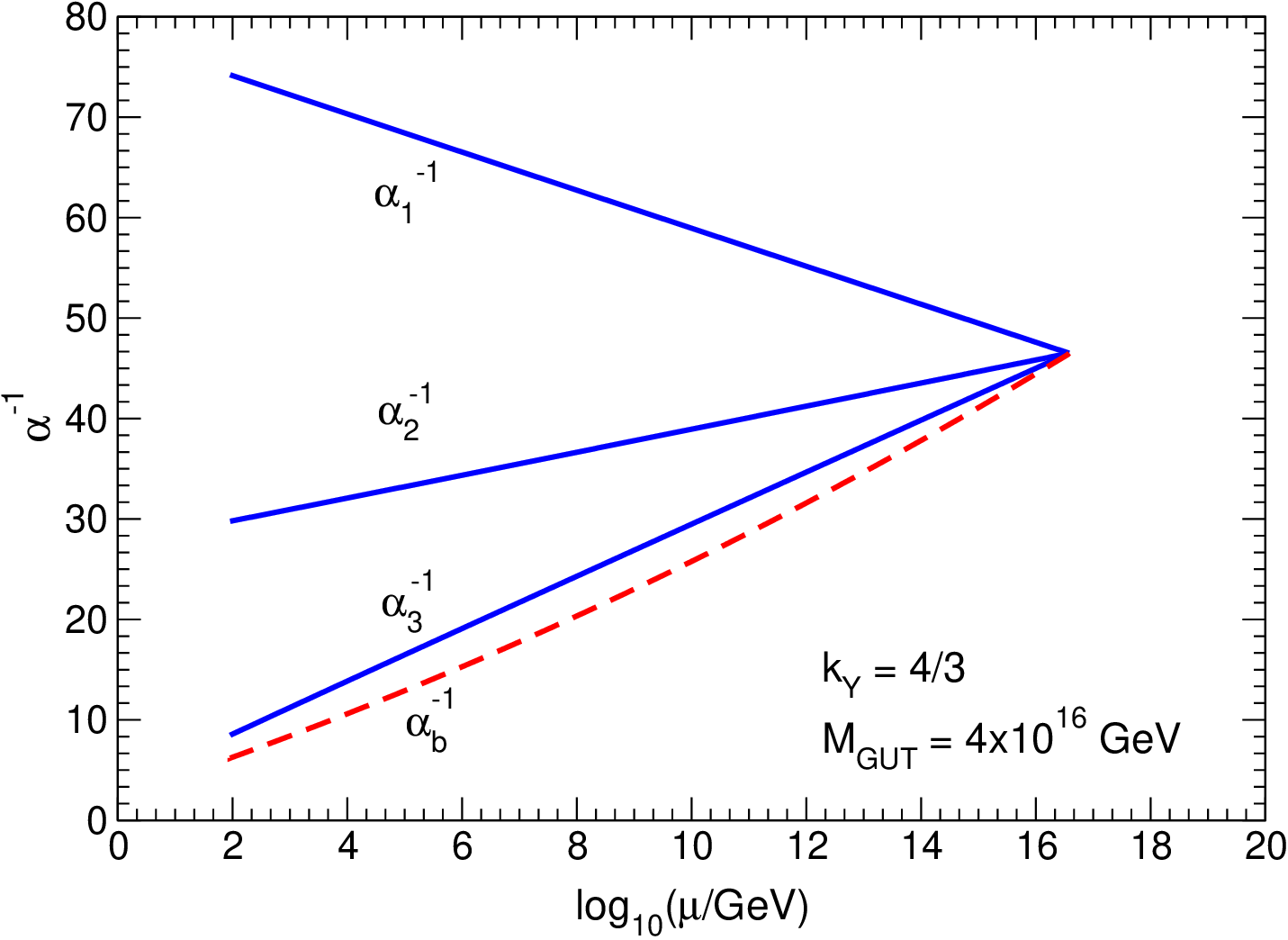}
\vspace{-0.3cm} \caption{\small  Two-loop evolution of gauge (solid)
and  bottom quark Yukawa (dash) couplings , with $k_Y=4/3$.}
\end{figure}

The corresponding charges $Qij$ are:
\begin{eqnarray}
&&Q12=\left (\mathbf{1\over 6}, \mathbf{-{\sqrt
{21}}}, \mathbf{2}\right ),~~~~~ 
Q13=\left (\mathbf{-{1\over 3}},
\mathbf{2{\sqrt {21}}}, \mathbf{2}\right ),
\nonumber \\
&&Q14= \left (\mathbf{{{\sqrt {21}}\over {6}}},
\mathbf{7}, \mathbf{1}\right ),~~~~~~~~~ Q34=\left (\mathbf{{2+{\sqrt
{21}}}\over 6}, \mathbf{7-2{\sqrt {21}}}, \mathbf{-1}\right),
\nonumber \\
&&Q24 =\left (\mathbf{-{1+{\sqrt {21}}}\over 6},
\mathbf{7+{\sqrt {21}}}, \mathbf{-1}\right ),~~~~~ Q23=\left (\mathbf{-{1\over
2}}, \mathbf{3{\sqrt {21}}}, \mathbf{0}\right ).
\end{eqnarray}

In Table \ref{Spectrum-DOWN-7D}, we present the zero modes from  the
chiral multiplets $\Sigma_1$, $\Sigma_2$ and $\Sigma_3$. We identify
them with the left-handed doublet   ($Q_3$), right-handed
bottom quark $b^c$,
 and  a pair of Higgs doublets $H_u$ and $H_d$. From
the trilinear term in the 7D bulk action in Eq. (\ref{action7}) we
obtain the bottom quark Yukawa coupling
\begin{eqnarray}
 \int d^7 x \left[ \int d^2 \theta \ g_7 Q_3 b^c H_d
 + h.c. \right].~\,
\end{eqnarray}
Thus, at $M_{\rm GUT}$ we have
\begin{eqnarray}
g_1=g_2=g_3=y_b=g_7/\sqrt{V}, \label{mm11}
\end{eqnarray}
where $y_b$ is the bottom quark Yukawa coupling to $H_d$. Then the bottom
quark Yukawa coupling to the SM Higgs boson is given by
\begin{eqnarray}
h_b=y_b\cos\beta=g_7\cos\beta /\sqrt{V}~.~\,
\label{un21}
\end{eqnarray}

\begin{figure}[t]
\centering
\includegraphics[width=8cm]{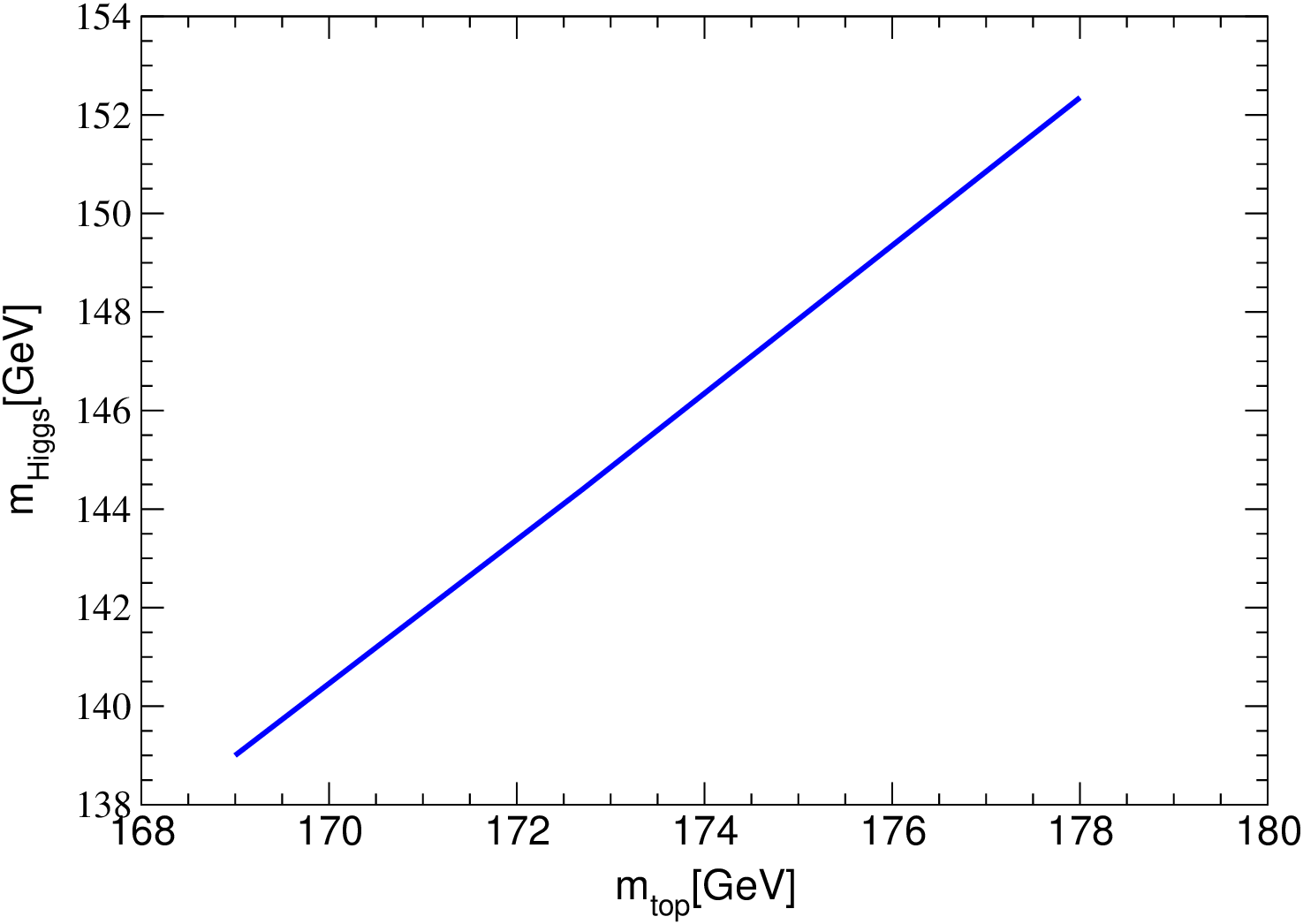}
\caption{\small Higgs boson mass versus $m_{top}$  with
gauge--bottom quark Yukawa coupling unification at $M_{\rm GUT}$.}
\end{figure}

Employing  the boundary  conditions from  Eq. (\ref{mm11}) and
proceeding analogously to the previous (top quark) case, we display
the four couplings in Fig.3. Using $m_b(m_b)=4.8$ GeV, we determine
the Higgs mass for this scenario to be $144  \pm 4$ GeV, as shown in
Fig. 4. The mixing angle $\beta$  is given by $\tan\beta\approx 82$,
very different from the value ($\tan\beta\approx1.5$) estimated in
the previous (top quark) Section.

\section{Gauge  and  Tau lepton Yukawa Coupling Unification}

To realize the gauge--tau lepton Yukawa coupling
unification, we set
\begin{eqnarray}
n_1~=~4~,~~n_2~=~3;~{\rm or}~ n_1~=~3~,~~n_2~=~2~.~\,
\label{N-numberB}
\end{eqnarray}
The generators for  $U(1)_Y \times U(1)_{\alpha} \times
U(1)_{\beta}$ are as follows:
\begin{eqnarray}
&&T_{U(1)_{Y}} \equiv  {1\over 2} ~ {\rm diag}\left(0, 0, 0, 0, 0,
1, -1 \right) - ~{{\sqrt {14}}\over {84}} {\rm diag}\left(4, 4, 4,
-3, -3, -3, -3 \right),~ \nonumber \\ && T_{U(1)_{\beta}} \equiv
-{{\sqrt {14}}\over {3}} ~{\rm diag}\left(0, 0, 0, 0, 0, 1, -1
\right) -{1\over 3} ~{\rm diag}\left(4, 4, 4, -3, -3, -3, -3
\right), \nonumber \\ &&T_{U(1)_{\alpha}} \equiv {\rm diag}\left(0,
0, 0, 1, 1, -1, -1 \right).
\end{eqnarray}
With ${\rm tr} [T_{U(1)_{Y}}^2]=2/3$, we obtain $k_Y =4/3$. This
insures the gauge coupling unification.

\renewcommand{\arraystretch}{1.4}
\begin{table}[b]
\begin{center}
\begin{tabular}{|c|c|}
\hline
Chiral Fields & Zero Modes  \\
\hline\hline
$\Sigma_1$ & $\tau^c$: $\mathbf{(1, 1)}_{Q34}$ \\
\hline $\Sigma_2$ & ~$H_d$:~ $\mathbf{(1, 2)}_{Q23}$;
~$H_u$:~ $\mathbf{(1, 2)}_{Q32}$ \\
\hline
$\Sigma_3$ & $L_3$: $\mathbf{(1, 2)}_{Q42}$ \\
\hline
\end{tabular}
\end{center}
\vspace{-0.3cm} \caption{ \small Zero modes from the chiral
multiplets $\Sigma_1$, $\Sigma_2$ and $\Sigma_3$   with gauge--tau
lepton Yukawa coupling unification. \label{Spectrum-TAU-7D}}
\end{table}

The $U(1)_Y \times U(1)_{\alpha} \times U(1)_{\beta}$ charges
$Qij$ are
\begin{eqnarray}
 &&Q12=\left (\mathbf{-{{\sqrt {14}}\over 12}}, \mathbf{-{{7}\over 3}}, \mathbf{-1}\right),~~~~~
Q13=\left (\mathbf{-{{6+\sqrt {14}}\over 12}},
\mathbf{-{{7-\sqrt {14}}\over 3}}, \mathbf{1}\right ), \nonumber \\
&&  Q23=\left (\mathbf{-{1\over 2}}, \mathbf{{\sqrt
{14}}\over {3}}, \mathbf{2}\right),~~~~~~~~~~  Q14=\left (\mathbf{{{6-\sqrt
{14}}\over 12}}, \mathbf{-{{7+\sqrt {14}}\over
3}}, \mathbf{1}\right),
\nonumber \\
&& Q24=\left (\mathbf{{1\over 2}}, \mathbf{-{{\sqrt
{14}}\over {3}}}, \mathbf{2}\right ),~~~~~~~~~~  Q34=\left (\mathbf{1},
 \mathbf{-{{2\sqrt {14}}\over {3}}}, \mathbf{0}\right ).
\end{eqnarray}
In Table \ref{Spectrum-TAU-7D}, we present the zero modes from  the
chiral multiplets $\Sigma_1$, $\Sigma_2$ and $\Sigma_3$. The zero
modes  include the third-family left-handed lepton doublet $L_3$,
one pair of Higgs doublets $H_u$ and $H_d$, and the right-handed
tau lepton $\tau^c$. From the trilinear term in the 7D bulk action, we obtain
the $\tau$ lepton Yukawa term
\begin{eqnarray}
 \int d^7 x \left[ \int d^2 \theta \ g_7 L_3 \tau^c H_d
 + h.c.\right].~\,
\end{eqnarray}
Thus, at  $M_{\rm GUT}$, we have
\begin{eqnarray}
g_1=g_2=g_3=y_{\tau}~,~\,
\end{eqnarray}
where  $y_{\tau}$ is the tau lepton Yukawa coupling.

This case turns out to be quite similar to the gauge--bottom quark
Yukawa coupling unification discussed above, with  $\tan\beta$ once
again large ($\sim$ 50 or so). The Higgs mass is   predicted to be
close to 144 GeV, with the usual uncertainty of several GeV arising
from the lack of a more precise determination of the top quark mass.

\section{$SU(8)$ Model}

It is possible to construct an $SU(8)$ model with $k_Y=4/3$, such
that the three SM gauge couplings as well as the two Yukawa
couplings are unified at $M_{\rm GUT}$, for example, the top and
bottom quark Yukawa couplings. From our previous discussions we note
that the unification of the gauge and top quark Yukawa couplings
favors a low value of $\tan \beta\sim1.5$, while the bottom quark
(or tau lepton) case requires a much larger value of $\tan
\beta\sim70-85$. Thus, we expect that a scenario in which all five
couplings are unified at $M_{\rm GUT}$ will lead to some
inconsistency. If we insist that the model correctly reproduces the
top quark mass, then the bottom quark mass will not be in agreement
with the data without invoking new physics such as
higher-dimensional operators. Mindful of this caveat the
construction of the $SU(8)$ model proceeds as follows. To break the
$SU(8)$ gauge symmetry, we choose the following $8\times 8$ matrix
representations for $R_{\Gamma_T}$ and $R_{\Gamma_S}$
\begin{eqnarray}
R_{\Gamma_T} &=& {\rm diag} \left(+1, +1, +1, \omega^{n_1},
\omega^{n_1}, \omega^{n_1}, \omega^{n_1}, \omega^{n_2} \right),~\,
\end{eqnarray}
\begin{eqnarray}
R_{\Gamma_S} &=& {\rm diag} \left(+1, +1, +1, +1, +1, -1, -1, -1
\right),~\,
\end{eqnarray}
where $n_1$ and $n_2$ are positive integers, and $n_1 \not= n_2$.
Then, we obtain
\begin{eqnarray}
 \{SU(8)/R_{\Gamma_T}\} ~=~ SU(3)_C\times SU(4)\times U(1) \times U(1)' ,~\,
\end{eqnarray}
\begin{eqnarray}
\{SU(8)/R_{\Gamma_S}\} ~=~ SU(5)\times SU(3) \times U(1),~\,
\end{eqnarray}
\begin{eqnarray}
 \{SU(8)/\{R_{\Gamma_T} \cup R_{\Gamma_S}\}\}
~=~ SU(3)_C\times SU(2)_L\times SU(2)_R \times U(1)_X \times
U(1)_{\alpha} \times U(1)_{\beta}.~\,
\end{eqnarray}
Therefore, we obtain that, for the zero modes, the 7D ${\cal N}=1$
supersymmetric $SU(8)$ gauge symmetry is broken down to the
4-dimensional ${\cal N}=1$ supersymmetric $SU(3)_C\times
SU(2)_L\times SU(2)_R \times U(1)_X \times U(1)_{\alpha} \times
U(1)_{\beta}$ gauge symmetry  \cite{Li:2001tx}.

We define the generators for the
$U(1)_X \times U(1)_{\alpha} \times U(1)_{\beta}$ gauge symmetry
 as follows
\begin{eqnarray}
T_{U(1)_{X}} &\equiv& {1\over 42} ~{\rm diag}\left(4, 4, 4, -3, -3,
-3, -3, 0 \right) + {{\sqrt {15}}\over {84}} ~T{\rm diag}\left(1, 1,
1, 1, 1, 1, 1, -7 \right),~\, \label{SU8-GUD1X} \nonumber
\\ T_{U(1)_{\alpha}} &\equiv& -{{\sqrt {15}}\over 3} ~{\rm
diag}\left(4, 4, 4, -3, -3, -3, -3, 0 \right) + ~{\rm diag}\left(1,
1, 1, 1, 1, 1, 1, -7 \right),~\, \nonumber \\
 T_{U(1)_{\beta}}
&\equiv& {\rm diag}\left(0, 0, 0, 1, 1, -1, -1, 0 \right).
\label{SU8-GUD1A}
\end{eqnarray}

The $SU(8)$ adjoint representation $\mathbf{63}$
is decomposed under the
$SU(3)_C\times SU(2)_L\times SU(2)_R \times U(1)_X \times U(1)_{\alpha}
\times U(1)_{\beta}$ gauge symmetry as
\begin{equation}
\mathbf{63} = \left(
\begin{array}{cccc}
\mathbf{(8,1, 1)}_{Q00} & \mathbf{(3, \bar 2, 1)}_{Q12}
& \mathbf{(3, 1, \bar 2)}_{Q13} & \mathbf{(3,1, 1)}_{Q14} \\
 \mathbf{(\bar 3,  2, 1)}_{Q21} & \mathbf{(1,3, 1)}_{Q00}
& \mathbf{(1, 2, \bar 2)}_{Q23} & \mathbf{(1, 2, 1)}_{Q24} \\
\mathbf{(\bar 3, 1, 2)}_{Q31} & \mathbf{(1, \bar 2, 2)}_{Q32}
& \mathbf{(1, 1, 3)}_{Q00} & \mathbf{(1, 1, 2)}_{Q34} \\
\mathbf{(\bar 3, 1, 1)}_{Q41} & \mathbf{(1, \bar 2, 1)}_{Q42}
& \mathbf{(1, 1, \bar 2)}_{Q43} & \mathbf{(1, 1, 1)}_{Q00}
\end{array}
\right) + 2 \mathbf{(1,1,1)}_{Q00}~,~\,
\end{equation}
where the  $\mathbf{(1,1,1)}_{Q00}$ in the fourth diagonal entry of
the matrix and the last term $2 \mathbf{(1,1,1)}_{Q00}$ denote the
gauge fields for the $U(1)_X \times U(1)_{\alpha} \times
U(1)_{\beta} $ gauge symmetry. Moreover,  the subscripts $Qij$,
which are anti-symmetric ($Qij=-Qji$), are the charges under the
$U(1)_X \times U(1)_{\alpha} \times U(1)_{\beta}$ gauge symmetry.
The subscript $Q00~=~ (\mathbf{0}, \mathbf{0}, \mathbf{0})$, and the
other subscripts $Qij$ with $i\not= j$ are

\begin{eqnarray}
&&Q12=\left (\mathbf{1\over 6}, \mathbf{-{{7{\sqrt
{15}}}\over {3}}}, \mathbf{-1}\right ),
~~~~~~~~~~~~~~~~~ Q13=\left
(\mathbf{1\over 6}, \mathbf{-{{7{\sqrt
{15}}}\over {3}}}, \mathbf{1}\right), \nonumber \\
 &&Q14=\left (\mathbf{{{2+2{\sqrt
{15}}}\over {21}}}, \mathbf{{{{24-4\sqrt {15}}}\over
{3}}}, \mathbf{0}\right ),~~~
 Q23=\left (\mathbf{0}, \mathbf{0}, \mathbf{2}\right),
\nonumber \\
 && Q24=\left (\mathbf{{{{-3+4\sqrt {15}}}\over {42}}},
\mathbf{8+\sqrt {15}}, \mathbf{1}\right ),\nonumber \\
 && Q34 = \left
(\mathbf{{{{-3+4\sqrt {15}}}\over {42}}},
\mathbf{8+\sqrt {15}}, \mathbf{-1}\right).
\end{eqnarray}

The $Z_6\times Z_2$ transformation properties for the decomposed components
of $V$, $\Sigma_1$, $\Sigma_2$,  and $\Sigma_3$ are still given by
 Eqs. (\ref{bb7})--(\ref{bb10}).
And we choose $n_1=5$ and $n_2=2~{\rm or}~3$, as in Eq.
(\ref{N-numberA}).

In Table \ref{Spectrum-SU8}, we present the zero modes from  the
chiral multiplets $\Sigma_1$, $\Sigma_2$ and $\Sigma_3$. The zero
modes include the left-handed quark doublet $Q_3$ for the third
family, one pair of  bidoublet Higgs fields $\Phi$ and
$\overline{\Phi}$, and the right-handed  quark doublet
$\overline{Q}_3$ for the third family.  More concretely, the
bidoublet Higgs field $\Phi$ contains a pair of Higgs doublets $H_u$
and $H_d$, and the right-handed  quark doublet $\overline{Q}_3$ for
the third family contains $t^c$ and $b^c$.

\renewcommand{\arraystretch}{1.4}
\begin{table}[t]
\begin{center}
\begin{tabular}{|c|c|}
\hline
Chiral Fields & Zero Modes  \\
\hline\hline
$\Sigma_1$ & $Q_3$:~ $\mathbf{(3, \bar 2, 1)}_{Q12}$ \\
\hline
$\Sigma_2$ &
~$\Phi$:~ $\mathbf{(1, 2, \bar 2)}_{Q23}$;
~$\overline{\Phi}$:~ $\mathbf{(1, \bar 2, 2)}_{Q32}$ \\
\hline
$\Sigma_3$ & $\overline{Q}_3$: $\mathbf{(\bar 3, 1, 2)}_{Q31}$ \\
\hline
\end{tabular}
\end{center}
\vspace{-0.3cm}
 \caption{\small The zero modes of the chiral multiplets
$\Sigma_1$, $\Sigma_2$ and $\Sigma_3$ in the 7D $SU(8)$  orbifold
model. \label{Spectrum-SU8}}
\end{table}

From the trilinear term in the 7D bulk action, we obtain the quark
Yukawa term
\begin{eqnarray}
\int d^7 x \left[ \int d^2 \theta \ g_8 Q_3 \overline{Q}_3 \Phi
 + h.c.\right],~\,
\end{eqnarray}
where $g_8$ is the $SU(8)$ gauge coupling at $M_{\rm GUT}$.

In order to break the $SU(2)_R\times U(1)_X$ gauge symmetry down to
the $U(1)_Y$ gauge symmetry, we introduce one pair of Higgs doublets
$H_1$ and $H_2$ with quantum numbers $\mathbf{( 2, -1/2)}$ and
$\mathbf{( 2, +1/2)}$ under the $SU(2)_R\times U(1)_X$ gauge
symmetry on the observable 3-brane, and assign the following VEVs:
\begin{eqnarray}
\langle H_1 \rangle ~=~\left(\begin{array}{c} v_X \\ 0
\end{array}\right) ,~~~~ \langle H_2 \rangle ~=~\left(\begin{array}{c}
0 \\ v_X \end{array}\right).~\,
\end{eqnarray}
The $U(1)_Y$ generator in $SU(8)$ is given by
\begin{eqnarray}
T_{U(1)_{Y}} &\equiv& {\rm diag}\left( {{8+{\sqrt {15}}}\over {84}},
{{8+{\sqrt {15}}}\over {84}}, {{8+{\sqrt {15}}}\over {84}},
{{-6+{\sqrt {15}}}\over {84}}, {{-6+{\sqrt {15}}}\over {84}},
\right.\nonumber\\&&\left. {{-48+{\sqrt {15}}}\over {84}},
{{36+{\sqrt {15}}}\over {84}}, -{{\sqrt {15}}\over {12}}\right).~\,
\label{SU8-GU1Y}
\end{eqnarray}
Because ${\rm tr} [T_{U(1)_{Y}}^2]=2/3$, we obtain $k_Y=4/3$.

With  $SU(2)_R\times U(1)_X$  broken  to  $U(1)_Y$, the third-family
quark Yukawa couplings are
\begin{eqnarray}
 \int d^7 x \left[ \int d^2 \theta \ g_8
\left( Q_3 t^c H_u +  Q_3 b^c H_d \right)
 + h.c.\right].~\,
\end{eqnarray}
 Thus, at the $M_{\rm GUT}$ scale, we have
\begin{eqnarray}
\label{ggg8} g_1=g_2=g_3=y_t=y_b\,.~\,
\end{eqnarray}

Employing the boundary conditions in Eq. (\ref{ggg8}) and making
sure that the top quark mass is reproduced correctly, we expect the
Higgs mass to be around $135\pm 6$ GeV. The bottom quark mass turns
out to be a factor two larger than its measured value and,  as
mentioned earlier, suitable non-renormalizable operators must be
introduced to rectify this. These additional operators are not
expected to significantly change the Higgs mass prediction.

\section{Conclusions}

We have considered a class of 7D orbifold GUTs with ${\cal N}=1$
supersymmetry in which the mass of the SM Higgs boson can be
reliably predicted. Depending on the details of the models the mass
is around 135 or 144 GeV, which is comfortably above the upper bound
on the mass of the lightest Higgs boson in the MSSM. The discovery
of the Higgs boson in the above mass range would be a boost for the
framework considered in this paper, namely that the unification of
the SM gauge couplings can be realized without low-energy
supersymmetry by invoking a non-canonical normalization of $U(1)_Y$.

\section*{Acknowledgments}
We are very grateful to Nefer Senoguz for pointing out a numerical
error in the previous version of the paper.  He also  kindly
provided us with new figures including two-loop RGEs for the Yukawa
and Higgs quartic  couplings. We would like to thank K. S. Babu, S.
M. Barr, S. Nandi, Z. Tavartkiladze and K. Tobe for discussions.
 T.L. would like to thank the Bartol
Research Institute for hospitality during the final stages of this
project. This work is supported in part by DOE Grant   \#
DE-FG02-84ER40163 (I.G.), \#DE-FG02-96ER40959 (T.L.) and   \#
DE-FG02-91ER40626 (Q.S.)


\section*{Appendix A: Seven-Dimensional Orbifold Models}

 We consider a 7D space-time $M^4\times T^2/Z_6 \times
S^1/Z_2$ with coordinates $x^{\mu}$, ($\mu = 0, 1, 2, 3$), $x^5$,
$x^6$ and $x^7$. The torus $T^2$ is homeomorphic to $S^1\times S^1$
and  the radii of the circles along the $x^5$, $x^6$ and $x^7$
directions are $R_1$, $R_2$, and $R'$, respectively. We define the
complex coordinate $z$ for $T^2$ and the real coordinate $y$ for
$S^1$,
\begin{eqnarray}
z \equiv{1\over 2} \left(x^5 + i x^6\right),~~~~~~~~~~ y \equiv x^7.
\end{eqnarray}
The torus $T^2$ can be defined by $C^1$ modulo the equivalent
classes:
\begin{eqnarray}
z \sim z+ \pi R_1 ,~~~~~~~~~~~ z \sim z +  \pi R_2 e^{{\rm
i}\theta}.
\end{eqnarray}
To obtain  the orbifold $T^2/Z_6$, we require that $R_1=R_2\equiv R$
and $\theta = \pi/3$. Then  $T^2/Z_6$  is obtained from $T^2$ by
moduloing the equivalent class
\begin{eqnarray}
\Gamma_T:~~~z \sim \omega  z,~\,
\end{eqnarray}
where $\omega =e^{{\rm i}\pi/3} $. There is one $Z_6$ fixed point
$z=0$, two $Z_3$ fixed points:
 $z=\pi R e^{{\rm i}\pi/6}/{\sqrt 3}$ and
$z=2 \pi R e^{{\rm i}\pi/6}/{\sqrt 3}$, and three $Z_2$ fixed
points: $z=\sqrt 3 \pi R e^{{\rm i}\pi/6}/2$, $z=\pi R/2$ and $z=
\pi R e^{{\rm i}\pi/3}/2$.
 The orbifold $S^1/Z_2$  is obtained from
$S^1$ by moduloing the equivalent class
\begin{eqnarray}
\Gamma_S:~~~y\sim -y~.~\,
\end{eqnarray}
There are two fixed points: $y=0$ and $y=\pi R'$.
The ${\cal N}=1$
supersymmetry in 7D has 16 supercharges corresponding to ${\cal
N}=4$ supersymmetry in 4D, and only the gauge multiplet can be
introduced in the bulk.  This multiplet can be decomposed under  4D
 ${\cal N}=1$ supersymmetry into a gauge vector
multiplet $V$ and three chiral multiplets $\Sigma_1$, $\Sigma_2$,
and $\Sigma_3$ in the adjoint representation, where the fifth and
sixth components of the gauge field, $A_5$ and $A_6$, are contained
in the lowest component of $\Sigma_1$, and the seventh component of
the gauge field $A_7$ is contained in the lowest component of
$\Sigma_2$.

We express the  bulk action in the Wess--Zumino gauge and 4D ${\cal
N}=1$ supersymmetry notation ~\cite{NMASWS}
\begin{eqnarray}
  {\cal S} &=& \int d^7 x \Biggl\{
  {\rm Tr} \Biggl[ \int d^2\theta \left( \frac{1}{4 k g^2}
  {\cal W}^\alpha {\cal W}_\alpha + \frac{1}{k g^2}
  \left( \Sigma_3 \partial_z \Sigma_2 + \Sigma_1 \partial_y \Sigma_3
   - \frac{1}{\sqrt{2}} \Sigma_1
  [\Sigma_2, \Sigma_3] \right) \right)
\nonumber\\
  &&
+ h.c. \Biggr]
 + \int d^4\theta \frac{1}{k g^2} {\rm Tr} \Biggl[
  (\sqrt{2} \partial_z^\dagger + \Sigma_1^\dagger) e^{-V}
  (-\sqrt{2} \partial_z + \Sigma_1) e^{V}
 + \partial_z^\dagger e^{-V} \partial_z e^{V}
\nonumber\\
  && +
  (\sqrt{2} \partial_y + \Sigma_2^\dagger) e^{-V}
  (-\sqrt{2} \partial_y + \Sigma_2) e^{V}
 + \partial_y e^{-V} \partial_y e^{V}
+ {\Sigma_3}^\dagger e^{-V} \Sigma_3 e^{V} \Biggr] \Biggr\},~\,
\label{action7}
\end{eqnarray}
where $k$ is the normalization of the group generator,  and ~${\cal
W_{\alpha}}$~ denotes the gauge field strength.  From the above
action, we obtain the transformations of the vector multiplet:
\begin{eqnarray}
  V(x^{\mu}, ~\omega z, ~\omega^{-1} {\bar z},~y) &=& R_{\Gamma_T}
 V(x^{\mu}, ~z, ~{\bar z},~y) R_{\Gamma_T}^{-1}~,~\,
\label{TVtrans}
\end{eqnarray}
\begin{eqnarray}
  \Sigma_1(x^{\mu}, ~\omega z, ~\omega^{-1} {\bar z},~y) &=&
\omega^{-1} R_{\Gamma_T} \Sigma_1(x^{\mu}, ~z, ~{\bar z},~y)
R_{\Gamma_T}^{-1}~,~\, \label{T1trans}
\end{eqnarray}
\begin{eqnarray}
   \Sigma_2(x^{\mu}, ~\omega z, ~\omega^{-1} {\bar z},~y) &=&
 R_{\Gamma_T}
\Sigma_2(x^{\mu}, ~z, ~{\bar z},~y)  R_{\Gamma_T}^{-1}~,~\,
\label{T2trans}
\end{eqnarray}
\begin{eqnarray}
 \Sigma_3(x^{\mu}, ~\omega z, ~\omega^{-1} {\bar z},~y)  &=&
\omega R_{\Gamma_T} \Sigma_3(x^{\mu}, ~z, ~{\bar z},~y)
R_{\Gamma_T}^{-1}~,~\, \label{T3trans}
\end{eqnarray}
\begin{eqnarray}
  V(x^{\mu}, ~z, ~ {\bar z},~-y) &=& R_{\Gamma_S}
 V(x^{\mu}, ~z, ~{\bar z},~y) R_{\Gamma_S}^{-1}~,~\,
\label{SVtrans}
\end{eqnarray}
\begin{eqnarray}
  \Sigma_1(x^{\mu}, ~ z, ~ {\bar z},~-y) &=&
 R_{\Gamma_S}
\Sigma_1(x^{\mu}, ~z, ~{\bar z},~y) R_{\Gamma_S}^{-1}~,~\,
\label{S1trans}
\end{eqnarray}
\begin{eqnarray}
   \Sigma_2(x^{\mu}, ~ z, ~ {\bar z},~-y) &=&
-  R_{\Gamma_S} \Sigma_2(x^{\mu}, ~z, ~{\bar z},~y)
R_{\Gamma_S}^{-1}~,~\, \label{S2trans}
\end{eqnarray}
\begin{eqnarray}
 \Sigma_3(x^{\mu}, ~ z, ~ {\bar z},~-y)  &=&
- R_{\Gamma_S} \Sigma_3(x^{\mu}, ~z, ~{\bar z},~y)
R_{\Gamma_S}^{-1}~,~\, \label{S3trans}
\end{eqnarray}
where we introduce  non-trivial transformation $R_{\Gamma_T}$ and
$R_{\Gamma_S}$ to break the bulk gauge group $G$.

The $Z_6\times Z_2$ transformation properties for the decomposed
components of $V$, $\Sigma_1$, $\Sigma_2$,  and $\Sigma_3$
in our $SU(7)$ and $SU(8)$ models are given by
\begin{equation}
V : \left(
\begin{array}{cccc}
(1, +) & (\omega^{-n_1}, +) & (\omega^{-n_1}, -) &
(\omega^{-n_2}, -)  \\
(\omega^{n_1}, +) & (1, +) & (1, -) & (\omega^{n_1-n_2}, -) \\
(\omega^{n_1}, -) & (1, -) & (1, +) & (\omega^{n_1-n_2}, +) \\
(\omega^{n_2}, -) & (\omega^{n_2-n_1}, -) & (\omega^{n_2-n_1}, +) &
(1, +)
\end{array}
\right)  +  (1, +) ~,~\, \label{bb7}
\end{equation}
\begin{equation}
\Sigma_1 : \left(
\begin{array}{cccc}
(\omega^{-1}, +) & (\omega^{-n_1-1}, +) & (\omega^{-n_1-1}, -) &
(\omega^{-n_2-1}, -)  \\
(\omega^{n_1-1}, +) & (\omega^{-1}, +) & (\omega^{-1}, -) & (\omega^{n_1-n_2-1}, -) \\
(\omega^{n_1-1}, -) & (\omega^{-1}, -) & (\omega^{-1}, +) & (\omega^{n_1-n_2-1}, +) \\
(\omega^{n_2-1}, -) & (\omega^{n_2-n_1-1}, -) & (\omega^{n_2-n_1-1},
+)  & (\omega^{-1}, +)
\end{array}
\right) +  (\omega^{-1}, +) ~,~\, \label{bb8}
\end{equation}
\begin{equation}
\Sigma_2 : \left(
\begin{array}{ccccc}
(1, -) & (\omega^{-n_1}, -) & (\omega^{-n_1}, +) &
(\omega^{-n_2}, +) \\
(\omega^{n_1}, -) & (1, -) & (1, +) & (\omega^{n_1-n_2}, +) \\
(\omega^{n_1}, +) & (1, +) & (1, -) & (\omega^{n_1-n_2}, -) \\
(\omega^{n_2}, +) & (\omega^{n_2-n_1}, +) & (\omega^{n_2-n_1}, -) &
(1, -)
\end{array}
\right) +  (1, -) ~,~\, \label{bb9}
\end{equation}
\begin{equation}
\Sigma_3 : \left(
\begin{array}{ccccc}
(\omega, -) & (\omega^{-n_1+1}, -) & (\omega^{-n_1+1}, +) &
(\omega^{-n_2+1}, +)  \\
(\omega^{n_1+1}, -) & (\omega, -) & (\omega, +) & (\omega^{n_1-n_2+1}, +) \\
(\omega^{n_1+1}, +) & (\omega, +) & (\omega, -) & (\omega^{n_1-n_2+1}, -) \\
(\omega^{n_2+1}, +) & (\omega^{n_2-n_1+1}, +) & (\omega^{n_2-n_1+1},
-) & (\omega, -)
\end{array}
\right) +  (\omega, -) ~,~\, \label{bb10}
\end{equation}
where the zero modes transform as $(1,+)$.

From Eqs. (\ref{bb7})--(\ref{bb10}), we find that  the 7D ${\cal N}
= 1 $ supersymmetric gauge symmetry $SU(7)$ and $SU(8)$ is broken down to  4D
${\cal N}=1$ supersymmetric gauge symmetry $SU(3)_C\times
SU(2)_L\times U(1)_Y \times U(1)_{\alpha} \times U(1)_{\beta} $
and $SU(3)_C\times
SU(2)_L\times SU(2)_R \times U(1)_X \times U(1)_{\alpha} \times U(1)_{\beta} $,
respectively \cite{Li:2001tx}. In addition, there are  zero modes from the chiral
multiplets $\Sigma_1$, $\Sigma_2$ and $\Sigma_3$ which play an
important role in gauge--Higgs--Yukawa unification.

\section*{Appendix B: Renormalization Group Equations}

The two-loop RGEs for the gauge couplings are \cite{mac}
\begin{equation}
(4\pi)^2\frac{d}{dt}~ g_i=g_i^3b_i +\frac{g_i^3}{(4\pi)^2} \left[
\sum_{j=1}^3 B_{ij}g_j^2-\sum_{\alpha=u,d,e} d_i^\alpha {\rm
Tr}\left( h_{\alpha}^{ \dagger}h_{\alpha}\right) \right],~\,
\label{2lgauge}
\end{equation}
The beta-function coefficients  for  $SU(3)_c\times SU(2)_L \times
U(1)_Y$ , with non-canonical $k_Y=\frac{4}{3}$ normalization for
$U(1)_Y$,  are

\begin{equation}
b_i=\left( -7,\, -\frac{19}{6},\, \frac{41}{8}\right),~~~~\,\,\,\,
b_{ij} = \left(
\begin{array}{ccc}
-26 & \frac{9}{2}
& \frac{11}{8}  \\
 12 & \frac{35}{6}
&  \frac{1}{2}\\
11 &\frac{27}{4} & \frac{199}{32}
\end{array}
\right),
\end{equation}

\begin{eqnarray}
&&d^u=\left(2 ,\frac{3}{2},\frac{17}{8} \right),~~~~~~
d^d=\left(2,\frac{3}{2},\frac{5}{8}\right),~~~~~~
d^e=\left(0,\frac{1}{2},\frac{15}{8}\right).~\,
\end{eqnarray}

The two-loop RGE for the Yukawa couplings and the Higgs quartic
coupling $\lambda$, with non-canonical $k_Y=\frac{4}{3}$
normalization for $U(1)_Y$, are
\begin{eqnarray}
\frac{d}{dt}~h_u&=&\frac{h_u}{16\pi^2}\left[ -\sum_{i=1}^3c_i^u
g_i^2 +\frac{3}{2} h_{u}^2 -\frac{3}{2} h_{d}^2
+\Delta_2\right]\nonumber \\ &+& \frac{h_u}{(16\pi^2)^2}\left[
\frac{1187}{384}g^4_1-\frac{23}{4}g^4_2-108g^4_3-\frac{9}{16}g^2_1g^2_2+
\frac{19}{12}g^2_1g^2_3+9g^2_2g^2_3+ \frac{5}{2}\Delta_3
\right.\nonumber\\   &+&\left. \left[
\frac{223}{64}g^2_1+\frac{135}{16}g^2_2+ 16 g_3^2\right] h_u^2
 - \left[
\frac{43}{64}g^2_1-\frac{9}{16}g^2_2+ 16 g_3^2\right] h_d^2
-6\lambda h_u^2 \right.\nonumber\\   &+&\left. \frac{3}{2}h_u^2-
\frac{5}{4}h_u^2h_d^2+\frac{11}{4}h_d^4+\left[\frac{5}{4}h_d^2
-\frac{9}{4}h_u^2 \right]\Delta_2 -\Delta_5
+\frac{3}{2}\lambda^2\right]~,~ \label{SMY3}
\end{eqnarray}

\begin{eqnarray}
\frac{d}{dt}~h_d&=&\frac{h_d}{16\pi^2}\left[ -\sum_{i=1}^3c_i^dg_i^2
-\frac{3}{2} h_u^{\dagger}h_{u} +\frac{3}{2} h_{d}^{\dagger}h_{d}
+\Delta_2 \right]
\nonumber\\
 &+& \frac{h_d}{(16\pi^2)^2}\left[ -\frac{127}{384}g_1^2-\frac{23}{4}g_2^2-108g_3^2
 -\frac{27}{16}g_1^2g_2^2+\frac{31}{12}g_1^2g_3^2+9g_2^2g_3^2
\right.\nonumber\\   &-&\left.
\left[\frac{79}{64}g_2^2-\frac{9}{16}g_2^2+16g_3^2 \right]h_u^2+
\left[\frac{187}{64}g_2^2+\frac{135}{16}g_2^2+16g_3^2
\right]h_d^2+\frac{5}{2}\Delta_3 \right.\nonumber\\   &-&\left.
6\lambda h_d^2 +\frac{3}{2}h_d^4-\frac{5}{4}h_d^2h_u^2
+\frac{11}{4}h_u^4+\left[ \frac{5}{4}h_u^2-\frac{9}{4}h_d^2
\right]\Delta_2+\Delta_5+\frac{3}{2}\lambda^2 \right],
     \label{SMY2}
\end{eqnarray}
\begin{eqnarray}
\frac{d}{dt}~h_e&=&\frac{h_e}{16\pi^2}\left[ -\sum_{i=1}^3c_i^eg_i^2
+\frac{3}{2} h_{e}^{\dagger} h_{e} +
\Delta_2 \right]\nonumber\\
 &+& \frac{h_d}{(16\pi^2)^2}\left[\frac{1371}{128}g_1^4 +\frac{23}{4}g_2^4
+\frac{27}{16}g_1^2g_2^2+
\left[\frac{387}{64}g_1^2+\frac{135}{16}g_2^2 \right]h_e^2
\right.\nonumber\\   &+&\left.
\frac{5}{2}\Delta_3-6\lambda_e^2+\frac{3}{2}h_e^4
+\frac{9}{4}\Delta_2h_e^2+\Delta_5+\frac{3}{2}\lambda^2
 \right],
 \label{SMY}
\end{eqnarray}

\begin{eqnarray}
\label{ggg7} \frac{d}{dt}~\lambda &=&\frac{1}{16\pi^2} \left[ 12
\lambda^2 -\left[\frac{9}{4} g_1^2 + 9 g_2^2 \right] \lambda
+{9\over 4} \left[ \frac{1}{3} \frac{3}{16} g_1^4 + \frac{1}{2}
g_1^2g_2^2 + g_2^4 \right]
 4\Delta_2 \lambda - 4 \Delta_4 \right] \nonumber\\ &+&
\frac{1}{(16\pi^2)^2}\left[\left[
\frac{27}{2}g_1^2+54g_2^2\right]\lambda^2 + \left[
\frac{73}{8}g_2^4+\frac{117}{16}g_1^2g_2^2+\frac{1887}{128}g_2^4
\right]\lambda +\frac{305}{8}g_2^6
\right.\nonumber\\
&-&\left. \frac{867}{96}g_1^2g_2^4-\frac{3411}{512}g_1^6
+64g_3^2\left[ h_u^4+h_d^4\right] -\frac{1}{2}g_1^2\left[
2h_u^4-h_d^4+3h_e^4 \right]
\right.\nonumber\\
&+&\left.
 \frac{3}{4}g_1^2\left[\left[
21g_2^2-\frac{57}{8}g_1^2\right]h_u^2+\left[
\frac{15}{8}g_1^2+9g_2^2 \right]h_d^2+\left[
11g_2-\frac{75}{8}g_1^2\right]h_e^2\right]
\right.\nonumber\\
&-&\left. \frac{3}{2}g_2^4\Delta_2
-\lambda\Delta_4+24\lambda^2\Delta_2+10\lambda\Delta_3-42\lambda
h_u^2h_d^2 +20\left[ 3h_u^6+3h_d^6+h_e^6\right]
\right.\nonumber\\
&+&\left. 12\left[h_u^4h_d^2+h_u^2h_d^4\right]
-78\lambda^3-\frac{1677}{128}g_1^4g_2^2\right],
\end{eqnarray}

where
\begin{eqnarray}
c^u_i=\left( 8 , \frac{9}{4},\frac{17}{16}  \right),
~~~~~~~c^d_i=\left( 8 , \frac{9}{4},\frac{5}{16} \right),
~~~~~~~c^e_i=\left(0 , \frac{9}{4}, \frac{45}{16} \right),~
\end{eqnarray}
\begin{eqnarray}
\Delta_2 =  3\,h_{u}^2+3\, h_{d}^2+ h_{e}^2\, ,
\end{eqnarray}
\begin{eqnarray}
\Delta_3=\sum c_i^u g_i h_u^2+\sum c_i^d g_i h_d^2+\frac{1}{3}\sum
c_i^eg_i^2h_e\, ,
\end{eqnarray}
\begin{eqnarray}
\Delta_4 =   3 \,h_{u}^4+3\, h_{d}^4 + h_{e}^4\, ,
\end{eqnarray}
\begin{eqnarray}
\Delta_5 = \frac{9}{4} \left[ 3 \,h_{u}^2+3\, h_{d}^2 +
h_{e}^2-\frac{2}{3}h_u^2h_d^2\right].
\end{eqnarray}

\end{document}